# Fiber Endoscopy Using Synthetic Wavelengths for 3D tissue imaging

Muralidhar Madabhushi Balaji[1,*], Partrick Cornwall[1], Parker Liu[1], Stefan Forschner[1], Jürgen Czarske[2], and Florian Willomitzer[1,*]

[1]*Wyant College of Optical Sciences, University of Arizona, United States of America*
[2]*Department of Electrical & Computer Engineering, TU Dresden, Germany*
[*]*mmuralidhar@arizona.edu, fwillomitzer@arizona.edu*

**Abstract:** Fiber-based endoscopes utilizing multi-core fiber (MCF) bundles offer the capability to image deep within the human body, making them well-suited for imaging applications in minimally invasive surgery or diagnosis. However, the optical fields relayed through each fiber core can be significantly affected by phase scrambling from height irregularities at the fiber ends or potential multi-mode cores. Moreover, obtaining high-quality endoscopic images commonly requires the fiber tip to be placed close to the target or relies on the addition of a lens. Additionally, imaging through scattering layers after the fiber tip is commonly not possible. In this work, we address these challenges by integrating Synthetic Wavelength Imaging (SWI) with fiber endoscopy. This novel approach enables the endoscopic acquisition of holographic information from objects obscured by scattering layers. The resulting endoscopic system eliminates the need for lenses and is inherently robust against phase scrambling caused by scattering and fiber bending. Using this technique, we successfully demonstrate the endoscopic imaging of $\approx 750 \mu m$ features on an object positioned behind a scattering layer. This advancement represents significant potential for enabling spatially resolved three-dimensional imaging of objects concealed beneath tissue using fiber endoscopes, expanding the capabilities of these systems for medical applications.

## 1. Introduction

Fiber endoscopes provide a minimally invasive solution for imaging in challenging environments where non-invasive methods are impractical [1]. In biomedical imaging, for example, they enable keyhole access to probe deep within internal organs that are otherwise inaccessible using traditional optical techniques. These capabilities have made fiber endoscopy increasingly valuable in applications such as surgical guidance or diagnostics. Typically, multi-core fiber (MCF) bundles or image guides are employed for spatially resolved imaging in these applications. The MCF bundles can be either positioned in close proximity to the sample of interest or combined with an attached lens to form an image [2]. However, both configurations face limitations in acquiring three-dimensional information and handling scattering effects, which are frequently encountered in clinical and other practical scenarios.

Recent efforts have explored the use of coherent imaging techniques, such as holography, for lens-less imaging with MCF bundles [2,3]. These approaches allow for lens-less imaging at larger working distances and enable the acquisition of phase information. However, they impose stringent requirements on the fiber core size and core-to-core spacing to mitigate phase scrambling while light propagates through the fiber. These constraints limit the adaptability of related approaches to different wavebands. Furthermore, traditional holographic methods face challenges in handling additional phase scrambling caused by scattering in scenarios where a scattering medium lies between the target object and the image guide [3]. Examples of such scenarios include spatially resolved measurement of blood flow variations beneath tissue during surgery or mapping the 3D structure of blood vessels.

In this work, we address the challenges of phase scrambling in image guides and multi-core multi-mode fibers (MC-MMF) using Synthetic Wavelength Imaging (SWI). Through SWI, we enable endoscopes to acquire holographic information of objects hidden behind additional scattering layers. SWI is an imaging technique originally designed to image hidden objects behind scattering media or in a Non-Line-of-Sight (NLoS) setting [4]. It has since been applied to a variety of scenarios, including single-shot 3D imaging, or phase measurements through multi-mode fibers [4-7]. In this contribution, we use a special flavor of SWI,

"Synthetic Wavelength Holography" (SWH), for lens-less endoscopic imaging through additional layers of scatter.

SWI leverages the existence of spectral correlations between optical fields acquired at two closely spaced optical wavelengths. These spectral correlations allow us to computationally construct a complex valued field at a larger *"synthetic wavelength"*. This *"synthetic field"* emulates most of the characteristics of an electromagnetic field operating at the larger *"synthetic wavelength"*. Due to the large synthetic wavelength, the multi-mode fiber cores act as single mode at the synthetic wavelength. Consequently, the approach is robust to phase scrambling in the individual fiber cores, as well as to bending the fiber. Moreover, the synthetic field demonstrates selective sensitivity to large-scale fluctuations introduced by the object/target while maintaining immunity to phase scrambling caused by scattering. We demonstrate that, despite significant scattering, our approach successfully recovers holographic information and reconstructs obscured objects. This holds significant potential for enabling spatially resolved 3D imaging of objects concealed beneath tissue using fiber endoscopes.

## 2. Method

In SWI, the object of interest is illuminated using two wavelengths $\lambda_1$ and $\lambda_2 = \lambda_1 + \Delta\lambda$, either directly (e.g. for 3D imaging [5]) or indirectly (e.g. NLoS imaging [4]). The two optical fields scattered off the object at these two wavelengths are recorded using standard digital holographic techniques. SWI leverages the fact that for a small wavelength separation $\Delta\lambda$, the two speckle fields remain correlated. The residual difference between these complex-valued optical fields encodes the information at the *"synthetic wavelength"*, defined as $\Lambda = \frac{\lambda_1 \lambda_2}{|\lambda_2 - \lambda_1|}$ [4]. Since $\Lambda$ can be chosen to be orders of magnitude larger than the optical wavelengths $\lambda_1$ and $\lambda_2$, the field at the synthetic wavelength is largely immune to the pathlength differences imparted by the scattering medium and retains the phase information of the object. Moreover, the significantly larger synthetic wavelength ensures that multi-mode fibers, which typically scramble light at optical wavelengths, behave effectively as single-mode fibers at $\Lambda$ (see Fig.1). Consequently, the synthetic wavelength holograms are immune to perturbations caused by multi-mode fiber and remain robust against fiber bending [7].

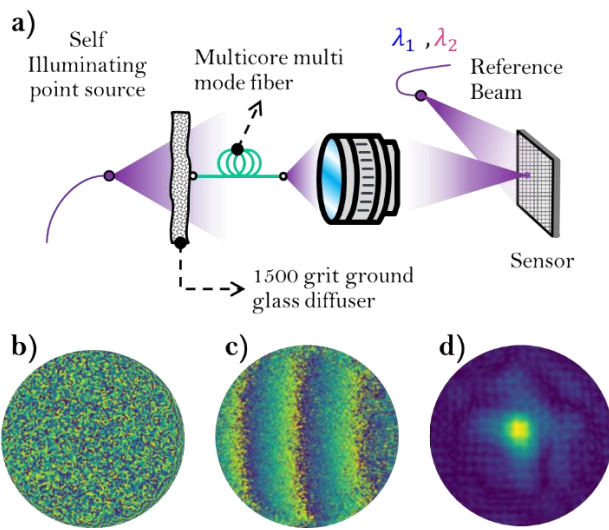

## 3. Results

We used the proof-of-principle setup depicted in Figure 1 to demonstrate the approach. In Fig.1a, a fiber tip, acting as a self-illuminating point source was placed behind a scattering medium (1500 grit ground glass diffuser). An image guide (MC-MMF, $\varnothing 3\ mm$ and cores of $\varnothing 50\ \mu m$), placed in contact with the scattering medium acts as a fiber endoscope. The light distribution from the scattering medium is recorded on a focal plane array that is imaging the distal end of the endoscope. The focal plane array records an off-axis hologram of the light distribution at two closely space wavelengths centered around $\lambda = 855\ nm$. While we have shown single-shot acquisition in [5], here, we record the two optical fields in a sequential manner for improved signal-to-noise performance.

Figure 1: Fiber Endoscopy Using Synthetic Wavelengths A) Setup schematic. The fiber time emulates a self-illuminated point object B) Phase at a single optical wavelength C) Phase at the synthetic wavelength and D) reconstructed target intensity.

The captured complex valued optical fields $E(\lambda_1)$ and $E(\lambda_2)$ are then computationally mixed to produce the field at the synthetic wavelength, i.e., $E(\Lambda) = E(\lambda_1)E^*(\lambda_2)$. It can be seen in Fig.1b that the phase map at the optical wavelength is completely randomized, thereby not encoding the positional information of the point source. This is due to the fact that the optical fields undergo significant phase scrambling while propagating through the scattering medium and the multi-mode fiber cores. However, the phase at the synthetic wavelength is not scrambled and exhibits curvature consistent with the presence of a point source (see Fig.1c). The object is reconstructed (including its lateral and depth position) by backpropagating the synthetic field in Fig.1c using angular spectrum propagation at the synthetic wavelength. The reconstruction of the point source is shown in Fig. 1c. The resolution of the reconstructed point source is consistent with the theoretically expected values (see below).

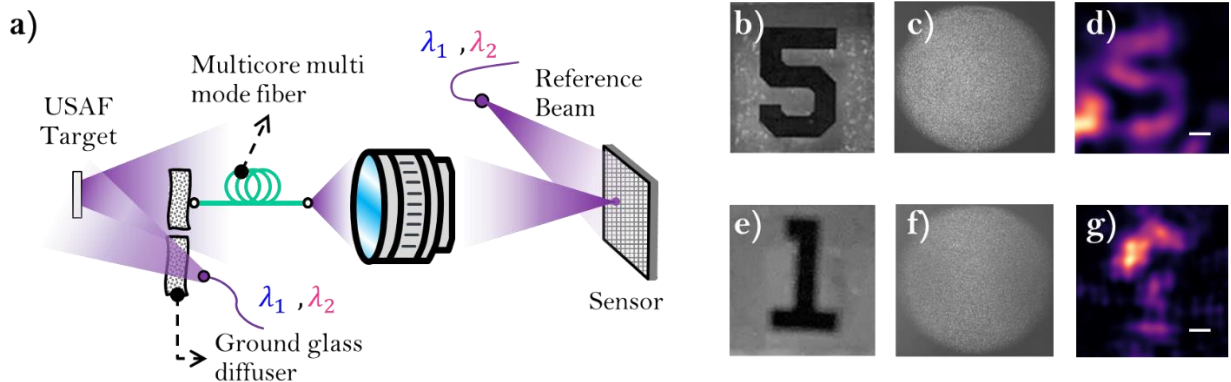

Figure 2: Fiber endoscopy of reflective targets embedded behind a scattering medium. a) setup schematic (b,e) intensity image of the reflective target (c,f) intensity image on the distal end of the endoscope (d,g) reconstructed image of the object at $\Lambda = 350\ \mu m$ using SWH. Scalebar: $750\mu m$.

In Fig. 2, we showcase the ability of SWI to endoscopically image extended objects hidden behind an additional scattering medium. Here, a USAF target was illuminated in reflection with a speckle field created by a diffuser (see Fig. 2a). The back-scattered light from the USAF target illuminates the other 1500-grit ground glass diffuser in contact with the MC-MMF. The USAF target was placed ~6 $mm$ behind the scattering medium. The synthetic wavelength hologram of the object behind the scattering medium was recorded at $\Lambda = 350$ μm and reconstructed using the procedure discussed above. The reconstructions of illuminated regions of the USAF target can be seen in Fig.2d and Fig.2g. It can be seen that the approach has the ability to recover $\approx 750\mu m$ features on the target (scalebar: $750\mu m$). This is consistent with the theoretical lateral resolution, defined as $\Lambda z/D$, where $z$ is the propagation distance and $D$ being the aperture diameter.

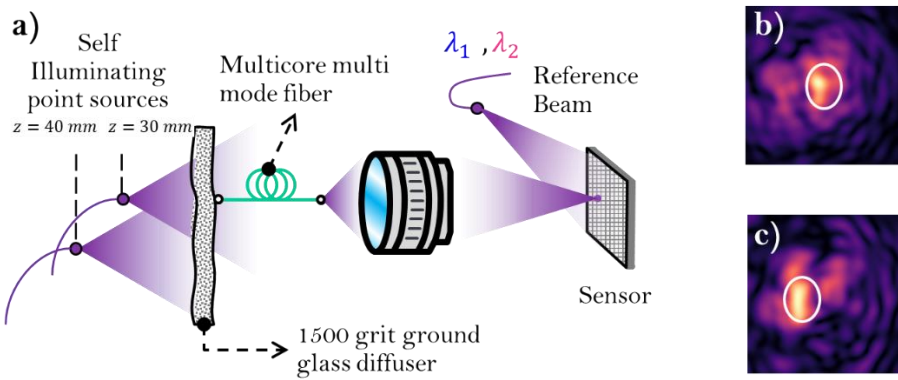

Figure 3: Localizing objects through scattering a) setup schematic (b,c) intensity image of the localized point sources at different depths..

Finally, we demonstrate the ability to localize multiple objects located at different depths behind scattering layers. Fig.3a depicts the schematic setup. Here, two fiber tips, acting as self-illuminating point sources were placed behind a scattering medium. Again, the MC-MMF of the earlier experiments was used to relay the scattered optical field to the sensor. To improve the depth separation between the reconstructed point sources, a series of synthetic fields were assembled over a frequency range of $1 THz$ with a frequency spacing of $100 GHz$. These fields were computationally interfered to generate a "synthetic pulse" of width 1 ps. The procedure of generating "synthetic pulses" is adopted from [4,6], and we refer to the respective publications for more information. The resulting reconstructions of both point sources at different depths can be seen in Fig.3b and c.

## 4. Discussion

The findings presented in this study highlight the capability of our method to endoscopically recover holographic information from objects obscured by scattering. The designed lens-less fiber endoscope system is robust to phase scrambling caused by scattering and fiber bending. Through this system we demonstrate the ability to image $\sim 750 \mu m$ features on an object positioned behind a scattering layer. This significantly enhances the performance and adaptability of fiber endoscopes in complex and challenging environments. Ongoing efforts are focused on implementing single-shot synthetic wavelength methods [5] to address dynamic scenarios, paving the way for real-time imaging in rapidly evolving conditions.

The presented work demonstrates significant potential for enabling spatially resolved 3D imaging of objects hidden beneath tissue using fiber endoscopes. For example, the ability to utilize larger core sizes and achieve spatially resolved imaging facilitates the acquisition of three-dimensional images of objects buried at greater depths. These advancements complement established techniques such as Optical Coherence Tomography (OCT) for tissue imaging. Collectively, these developments represent a promising step forward in advancing biomedical diagnostics and improving imaging performance in challenging settings.

**References:**

1. Lee, Cameron M., Christoph J. Engelbrecht, Timothy D. Soper, Fritjof Helmchen, and Eric J. Seibel. "Scanning fiber endoscopy with highly flexible, 1 mm catheterscopes for wide-field, full-color imaging." Journal of biophotonics 3, no. 5-6 (2010): 385-407.
2. Badt, Noam, and Ori Katz. "Real-time holographic lensless micro-endoscopy through flexible fibers via fiber bundle distal holography." *Nature Communications* 13.1 (2022): 6055.
3. Sun, Jiawei, Jiachen Wu, Song Wu, Ruchi Goswami, Salvatore Girardo, Liangcai Cao, Jochen Guck, Nektarios Koukourakis, and Jürgen W. Czarske. "Quantitative phase imaging through an ultra-thin lensless fiber endoscope." Light: Science & Applications 11, no. 1 (2022): 204.
4. Willomitzer, Florian, Prasanna V. Rangarajan, Fengqiang Li, Muralidhar M. Balaji, Marc P. Christensen, and Oliver Cossairt. "Fast non-line-of-sight imaging with high-resolution and wide field of view using synthetic wavelength holography." Nature communications 12, no. 1 (2021): 6647.
5. Ballester, Manuel, Heming Wang, Jiren Li, Oliver Cossairt, and Florian Willomitzer. "Single-shot synthetic wavelength imaging: Sub-mm precision ToF sensing with conventional CMOS sensors." Optics and Lasers in Engineering 178 (2024): 108165.
6. Cornwall, Patrick, Manuel Ballester, Stefan Forschner, Muralidhar Madabhushi Balaji, Aggelos Katsaggelos, and Florian Willomitzer. "Synthetic Light-in-Flight." arXiv preprint arXiv:2407.07872 (2024).
7. Forschner, Stefan, Patrick Cornwall, Manuel Ballester, Muralidhar Madabhushi Balaji, Jürgen Czarske, and Florian Willomitzer. "Towards Synthetic Wavelength Imaging through Multi-mode Fibers." In Proc. of SPIE Vol, vol. 13258, pp. 1325801-212. 2024.